\documentstyle[epsf,twocolumn]{jpsj}

\def\JH{J_{\rm H}}
\def\nmag{n_{\rm mag}}
\def\Ds{D_{\rm s}}

\catcode`\@=11
\def\braket#1{\left\langle#1\right\rangle}
\def\simle{\mathrel{\mathpalette\@versim<}}   % < over \sim
\def\simge{\mathrel{\mathpalette\@versim>}}   % > over \sim
\def\@versim#1#2{\lower2.5pt\vbox{\baselineskip0pt \lineskip-.5pt
   \ialign{$\m@th#1\hfil##\hfil$\crcr#2\crcr\sim\crcr}}}

\def\rmi{{\rm i}}

\def\Im{{\cal I}\!{\sl m}\,}
\catcode`\@=12
\title{
Unconventional One-Magnon Scattering Resistivity in Half-Metals
}
\author{Nobuo {\sc Furukawa}}

\def\runtitle{One-Magnon Scattering Resistivity in Half-Metals}

\def\runauthor{Nobuo {\sc Furukawa}}

\inst
{
Department of Physics, Aoyama Gakuin University, 
Setagaya, Tokyo 157-8572, Japan
}

\recdate
{
}

\abst
{
Low-temperature resistivity of  half-metals
is investigated. To date it has been discussed that
the one-magnon scattering process in half-metals
 is irrelevant for low-temperature
resistivity, due to the fully spin-polarized electronic structure
at the ground state. If one takes into account the non-rigid-band 
behavior of the minority band due to spin fluctuations 
at finite temperatures, however,
the unconventional one-magnon scattering process is shown to be most relevant
and gives $T^3$ dependence in resistivity.
This behavior may be used as a crucial test in the search for
half-metallic materials which are potentially important for applications.
Comparison with resistivity data of
 La$_{1-x}$Sr$_{x}$MnO$_3$ as candidates for half-metals
 shows  good agreement.
}

\kword
{
half-metal, ferromagnetic metal, one-magnon resistivity, 
colossal magnetoresistance manganites
}

\begin{document}
\sloppy
\maketitle

Since Zener\cite{Zener51} introduced the double exchange (DE) model 
as a model for perovskite manganites $A$MnO$_3$,
intensive studies have been performed.
Nevertheless, spin fluctuation effects beyond
the mean-field picture have not yet been fully clarified.
Spin fluctuations strongly influence the
structure of the  conduction electron band in the DE model.
Therefore, many-body treatments with respect to spin
fluctuations are important \cite{Furukawa99}.
One example of the mean-field treatment giving an inaccurate result
is that it substantially
overestimates  the Curie temperature $T_{\rm c}$.
Using  realistic parameters
for  La$_{1-x}$Sr$_{x}$MnO$_3$,
Millis {\em et al.}  \cite{Millis95} 
demonstrated that the mean-field treatment of the DE model gives
 $T_{\rm c} = 1000\ {\rm K} \sim 3000\ {\rm K}$,
 which is much larger than the experimental values
$T_{\rm c} = 300\ {\rm K} \sim 400\ {\rm K}$ at $0.2\simle x\simle 0.5$,
and concluded that 
the DE model alone cannot explain La$_{1-x}$Sr$_{x}$MnO$_3$.
However, by considering
thermal spin fluctuations at $T\sim T_{\rm c}$, which
reduce electronic hoppings, the model reproduces
experimental $T_{\rm c}$  \cite{Furukawa95b,Varma96}.

Resistivity is also strongly influenced
by  spin fluctuations.
Kubo and Ohata \cite{Kubo72} investigated the
resistivity of the DE model. 
The ground state
of this model is perfectly spin-polarized,
and is now called half-metallic\cite{Irkhin94}.
Based on a rigid-band picture,
they have proposed that 
two-magnon processes which give $\rho\propto T^{4.5}$
are  relevant for the low-temperature resistivity.
In this Letter, however,
we demonstrate that  a different scattering process creates
 $\rho\propto T^{3}$ in half-metals.
This process appears only by
considering spin fluctuations beyond the rigid-band
approximation.

Half-metals interest us 
from the viewpoint of fabricating a tunneling magnetoresistance junction,
since the perfect spin polarization of conduction electrons is
important for the sensitivity of  devices.
Indeed, the origin of
low-field magnetoresistance in polycrystal and trilayer junctions
of manganites is considered to be the
spin-dependent tunneling at half-metallic junctions.\cite{Hwang96,Sun96}
We also note that an exotic type of superconductivity in half-metals has been
predicted\cite{Pickett97a}.
In order to conduct a material search for half-metals,
a unique $T$-dependence in resistivity 
might play an important role as a crucial test.

\begin{figure}
\epsfxsize=3cm
\hfil\epsfbox{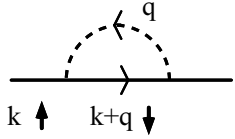}\hfil
\caption{1MS self-energy diagram for the majority band. The solid line
and the dashed curve represent electron and magnon Green's functions,
respectively.}
\label{FigSE}
\end{figure}

The magnetic scattering of quasiparticles is one of the
origins of resistivity in magnetic metals.
In conventional  itinerant weak ferromagnets,
the one-magnon scattering (1MS) process,\cite{Kasuya59,Mannari59}
illustrated in Fig.~\ref{FigSE},
 is one of the origins of
 resistivity in the form $\rho\propto T^2$.
The existence of coherent quasiparticles
for both majority- and minority-spin bands is
essential for the conventional 1MS process,
since the process involves spin-flipping vertices and a propagator
of the minority band at the Fermi level.

\begin{figure}[hbt]
\epsfxsize=8cm
\hfil\epsfbox{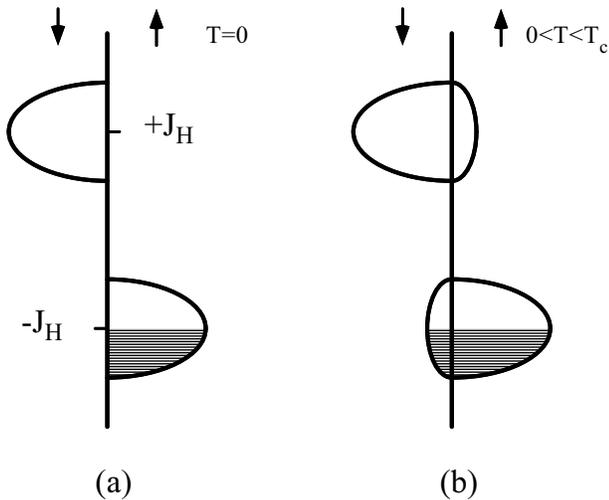}\hfil
\caption[DOS]
{DOS structure of the DE model at (a) $T=0$ and (b) $0<T<T_{\rm c}$,
where $\JH$ is Hund's coupling.}
\label{FigDOS}
\end{figure}

On the other hand, 
half-metals  belong to a different class of 
itinerant ferromagnets.
Conduction electrons are perfectly spin-polarized, and
the  Fermi surface is absent in the minority-spin band.
In Fig.~\ref{FigDOS}(a)
we illustrate the density of states (DOS)  for a large Hund's coupling ($\JH$) region of
the DE model as a canonical example.
Kubo and Ohata \cite{Kubo72} stated that
the 1MS is forbidden in the DE model
due to its half-metallic ground state.

However, 
spin fluctuations at a finite temperature
create modifications of the electronic band
structure. In Fig.~\ref{FigDOS}(b) we illustrate
the DOS at $0<T<T_{\rm c}$. 
 The DOS has been
calculated by the dynamical mean-field (DMF) theory
which takes  local spin fluctuations into account,
and also by the Monte Carlo method which treats spin fluctuations
on a finite size cluster in an controlled manner\cite{Furukawa99}.
At a finite temperature, spin fluctuations induce a minority band.
Once the thermally activated minority band is created and occupied,
a 1MS is allowed. 
This 1MS is  unconventional in the sense that
it is absent at the ground state and is  strongly 
affected by the spin fluctuation.

Based on such a non-rigid-band picture,
we calculate the resistivity of half-metals
in three dimensions.
We consider a magnon-electron interaction in the form
\begin{equation}
  {\cal H}_{\rm int} = \frac{g}{\sqrt N} \sum_{qk}\left( 
	      a_q^\dagger c_{k\uparrow}^\dagger c_{k+q\downarrow}
          +   a_q  c_{k+q\downarrow}^\dagger c_{k\uparrow}
	      \right),
\end{equation}
where $c^\dagger$ and $a^\dagger$ are the creation operators for
electrons and magnons, respectively. We define the quantization
axis in such a way that a magnon carries $S_z = -1$.
For the DE model,
the electron-magnon coupling constant is given by
$g = J_{\rm H}/ \sqrt{S}$ where $\JH$ gives the
Hund's coupling between conduction electrons
 and localized spins.\cite{Furukawa96} 
Therefore, a perturbational approach based on the linear magnon approximation
is well defined, at least in the limit $S\to\infty$, while $J_{\rm H}$ is kept
constant, and presumably gives  qualitatively correct results
at low temperatures where spin fluctuations are sufficiently small.

Since conductivity is governed by the majority carriers,
the self-energy for the majority spin electrons $\Sigma_\uparrow$
 determines the low-temperature resistivity.
A 1MS self-energy, illustrated in Fig.~\ref{FigSE},
 is given in the form
\begin{equation}
% && 
\Sigma_\uparrow(k,\omega) =
%\nonumber\\
% && \qquad 
  \frac{g^2}N\sum_{q} \int \frac{{\rm d}\omega'}{2\pi} 
  G_\downarrow(k+q,\omega+\omega')D(q,\omega'),
 \label{defSigmaOneMagnon}
\end{equation}
where $G_\downarrow$ and $D$ are Green's functions for
minority-spin electrons and magnons, respectively.
Note the direction of the magnon propagator.
The 1MS for a majority band electron occurs only
when it absorbs a magnon carrying $S_z=-1$. The intermediate state
involves a propagator of the {\em preoccupied} minority band.
Thus, the 1MS self-energy strongly depends on  spin fluctuations
and diminishes at $T\to0$.

Let us  assume that Green's function  for the minority band
obtained by the DMF \cite{Furukawa99} is
also valid  in  three-dimensional systems,
which at small frequencies reads 
\begin{equation}
 G_\downarrow(k,\omega) \simeq \frac{z_\downarrow}
  {\omega - \zeta_\downarrow(k) - \rmi \Gamma_\downarrow}.
  \label{defGdown}
\end{equation}
The quasiparticle renormalization factor scales as
\begin{equation}
 z_\downarrow = \delta m / 2,
\end{equation}
where $\delta m = (M(0)-M(T))/M(0)$ is the reduction of
the spin moment $M(T)$
scaled by its saturated value $M(0)$.
Quasiparticle dispersion relation is denoted by
$\zeta_\downarrow(k)  = (W^*/W) \varepsilon_k -\mu $.
Here, 
$\varepsilon_k$ is the dispersion relation
of the majority band at $T=0$, and $\mu$ is the chemical potential,
while
$W$ and $W^*$ are the quasiparticle bandwidth at 
$T=0$ and finite $T$, respectively,
which scales as $W^*/W \simeq \delta m/2$.
The inverse lifetime for the minority band due to
spin-disorder localization is given by
 $\Gamma_\downarrow
 \sim W (1-\delta m/2)$.

At low temperatures where $\delta m \ll 1$, we have
the incoherent limit for $G_\downarrow$, namely
$|\zeta_\downarrow(k)| \ll \Gamma_\downarrow$. Then, the conventional
1MS calculation becomes invalid. The integration
in eq.~(\ref{defSigmaOneMagnon}) is dominated by the pole in
the magnon Green's function. As a rough estimate, we have
\begin{eqnarray}
 \Im\Sigma_\uparrow &\sim& \frac{z_\downarrow}{\Gamma_\downarrow} 
  \frac1N\sum_{q,n} D(q,\rmi \Omega_n)
  \propto z_\downarrow  \braket{\nmag} ,
%     \nonumber\\
%  &\propto& z_\downarrow  \braket{\nmag} \sim (\delta m)^2
\end{eqnarray}
where $\braket{\nmag}$ is the magnon occupation number per site
which satisfies $ \delta m = \braket{\nmag}/ M(0)$.
Therefore, the inverse lifetime of the majority spin carrier at the Fermi level
$\Gamma_\uparrow = \Im\Sigma_{\uparrow}(k=k_{\rm F},\omega=0)$
is proportional to the square of the spin fluctuation,
$\Gamma_\uparrow \propto (\delta m)^2$.
For three-dimensional  magnons with the  dispersion
relation $ \omega_q = \Ds q^2$, where $\Ds$ is the spin stiffness,
we have $\braket{\nmag} \simeq 0.06 (T/\Ds)^{3/2}$.
Using the Drude formula
$\rho = (m^* / n e^2) \Gamma_\uparrow$,
 we obtain
\begin{equation}
  \rho  \propto (T/\Ds)^3.
  \label{rhoT3}
\end{equation}

More formal expression for $\Sigma_\uparrow$ is obtained through the
spectral representation as
\begin{eqnarray}
&&\Im \Sigma_{\uparrow}(k,\omega-\rmi \eta) = 
 \frac{g^2}{\pi N} \sum_q \int^\infty_{-\infty} \!\!\!{\rm d} \omega'
  \left[ \vphantom{\sum} f(\omega+\omega') + n(\omega')\right]
 \nonumber\\
&&\qquad\times\  
\Im G_\downarrow(k+q, \omega+\omega'-\rmi\eta)\,\, \Im D(q,\omega'-\rmi\eta),
  \label{SigmaSpect}
\end{eqnarray}
where $f(\omega)$ and $n(\omega)$ are Fermi and Bose distribution
functions, respectively.
The incoherent limit of  eq.~(\ref{defGdown}) gives
%\begin{equation}
$ \Im G_{\downarrow}(k,\omega-\rmi \eta) 
\simeq z_{\downarrow} / \Gamma_{\downarrow}$,
%\end{equation}
and for the magnon Green's function we use
%\begin{equation}
$ \Im D(q,\omega-\rmi \eta) = \pi \delta(\omega-\omega_q)$.
%\end{equation}
Then, we obtain
\begin{eqnarray}
 \Gamma_\uparrow &=&  g^2 \frac{z_\downarrow}{\Gamma_\downarrow}
 \frac1N\sum_q \left[ f(\omega_q) + n(\omega_q) \right]
\sim  g^2 \frac{z_\downarrow}{\Gamma_\downarrow} \braket{ n_{\rm mag}}.
\end{eqnarray}
Here, we use $f(\omega) \ll n(\omega)$ at small frequencies
which mostly contributes to the integration.

We point out here that
$T^3$ resistivity  can be used as a probe to investigate
 whether a given compound is a half-metal.
For example, if there exists a substantial overlap with other bands, 
it will create a conventional $T^2$ resistivity.
Wang and Zhang  calculated the case of  a nearly half-metal
where the slightly doped minority band is   Anderson-localized
below the mobility edge\cite{Wang99}. They derived
the resistivity $\rho\propto T^{2.5}$ at high temperatures which crosses over
to $\rho\propto T^{1.5}$ at low temperatures.

Let us now compare the result with those of experiments.
We investigate the resistivity data for
La$_{1-x}$Sr$_x$MnO$_3$ as a candidate for a half-metal.
It has previously been shown by various experiments
 that the spin polarization of the conduction
electron is very large;
large values of saturation moment $\sim 4\mu_{\rm B}$ suggest
that Mn ions are in  high-spin states.\cite{Jonker50a}
Spin-dependent photoemission experiments\cite{Park98a}
as well as tunneling magnetoresistance measurements\cite{Sun96}
indicate the absence of the minority-spin quasiparticles at the
Fermi level. Observation of  magnon dispersion throughout the
Brillouin zone\cite{Perring96} supports the hypothesis that the
low-energy Stoner continuum is absent
 due to large spin polarization.\cite{Furukawa96}
Nevertheless, since there exist overlaps of Mn $e_{\rm g}$ orbitals and
O $2p$ orbitals in the conduction band which might break
down the perfect spin-polarization, 
it is important to determine
whether these series of compounds are  half-metals or not,
Here, we examine whether they can be considered as perfectly spin-polarized
  half-metals, on the basis f the low-temperature resistivity measurements.

\begin{figure}
\epsfxsize=8cm
\hfil\epsfbox{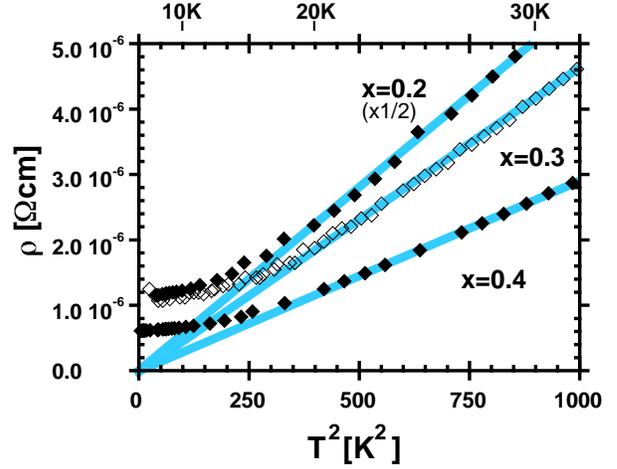}\hfil
\caption{
$T^2$-plot of the low-temperature resistivity of
La$_{1-x}$Sr$_x$MnO$_3$ at $x=0.2$, $0.3$ and $0.4$.
In order to fit all the data in a plot, origins  are shifted
for each $x$,  and $x=0.2$ data are plotted by 1/2 scale.
Grey lines represent $T^2$-fits for the high-temperature region.
}
\label{FigT2}
\end{figure}

Low-temperature resistivity data for 
single crystals of  La$_{1-x}$Sr$_x$MnO$_3$
are obtained from ref.~\citen{Urushibara95}. 
The residual resistivity of the sample is $\rho_0
 \simle 35 \mu\Omega{\rm cm}$
at $x=0.4$ indicating the high quality of the sample.
In Fig.~\ref{FigT2} we
show the conventional $T^2$-plot of the resistivity $\rho(T)$.
In the high-temperature region, the resistivity data fit well in the form
$\rho(T) = R_0 + A  T^\alpha$ with $\alpha \simeq 2$,
 as has been commonly  reported
for metallic manganites.\cite{Urushibara95,Schiffer95,Snyder96,Jaime98}

However, in the low-temperature region
below 20 {\rm K} we see substantial deviation from the $T^2$-like behavior.
The flattening of $\rho(T)$ is observed.
Moreover, there exist discrepancies between
$\rho_0 = \rho(T\to 0)$ and its counterpart by extrapolation
 from the high-temperature  $T^2$ region, $R_0$.
The reduction of the resistivity $R_0 < \rho_0$ indicates
that $\rho(T)$ is not understood by Matthiessen's rule,
{\em i.e.}, $\rho(T)$ at $T\simge20\ {\rm K}$ is not explained by
the simple sum of the residual scattering and the
Fermi-liquid-type scattering at $T\to 0$.
Namely, this indicates that not only the quasiparticle lifetime but also
the nature of the conduction channel itself might possibly make a crossover
at $T\sim 20\ {\rm K}$.

\begin{figure}
\epsfxsize=8cm
\hfil\epsfbox{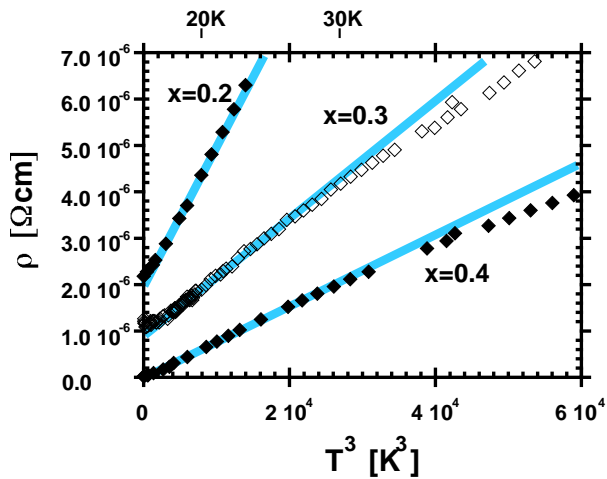}\hfil
\caption{
$T^3$-plot of the low-temperature resistivity of
La$_{1-x}$Sr$_x$MnO$_3$ at $x=0.2$, $0.3$ and $0.4$.
Origins are shifted for each $x$.
Grey lines represent $T^3$-fits for the low-temperature region.}
\label{FigT3}
\end{figure}

\begin{figure}
\epsfxsize=8cm
\hfil\epsfbox{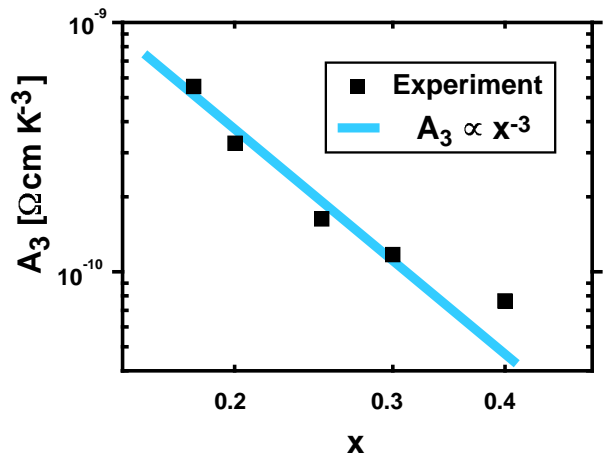}\hfil
\caption{Coefficient for the $T^3$ term of the resistivity, $A_3$, where
 $\rho(T)-\rho(0) = A_3 T^3$,
for La$_{1-x}$Sr$_x$MnO$_3$ at various $x$.
The line represents the least-square fit for $A_3 \propto x^{-3}$. }
\label{FigA3}
\end{figure}

Let us now restrict ourselves to the
low-temperature region below the crossover, and examine the 1MS results.
In Fig.~\ref{FigT3} we show a $T^3$-plot of the resistivity $\rho(T)$.
The data at $T\simle 30\ {\rm K}$ is
well  reproduced by
the unconventional 1MS contribution $\rho(T)=\rho_0 + A_3 T^3$.
We also investigate the coefficient of the $T^3$ term as a crucial test.
From eq.~(\ref{rhoT3}) we expect the scaling relation $A_3 \propto \Ds^{-3}$.
Unfortunately, the values of $\Ds$ have been measured for only a few
limited concentrations thus far, 
so a direct comparison cannot be carried out.
Instead, we assume
a scaling relating $\Ds\propto x$, which is experimentally
observed in La$_{1-x}$Sr$_{x}$MnO$_3$
 at $0.15 \simle x \simle 0.3$.\cite{Endoh97} 
In Fig.~\ref{FigA3} we plot the coefficient of the $T^3$ term
as a function of doping, and we roughly see $A_3 \propto x^{-3}$.
The deviation from the fit at $x=0.4$ can be understood by the
saturation of $\Ds$ as a function of $x$, since
we experimentally
 see that at $x\sim 0.4$ 
the increase of $T_{\rm c}$ by increasing $x$ saturates
 \cite{Urushibara95} and $\Ds$ roughly scales as
$T_{\rm c}$ \cite{Endoh97}.
In this analysis we assumed that the carrier number $n$ and
the effective mass $m^*$ do not change substantially
upon doping in the metallic region,
based on the Hall coefficient and 
specific-heat measurements\cite{Asamitsu98,Okuda98}.

The $T^3$ behavior as well as the concentration dependence
of its coefficient shows consistent
 theoretical and experimental results.
Another possible test 
is to measure the resistivity under high magnetic fields which
suppress $\delta m$. 
Resistivity due to  the unconventional 1MS
should scale as $\rho(T,H)-\rho_0 \propto (\delta m(T,H))^2$.

Thus, the low-temperature $T^3$
behavior in the resistivity of (La,Sr)MnO$_3$ 
is  direct evidence of their half-metallic nature.
Similar behaviors have been reported
for various compositions of doped manganites in the
ferromagnetic metal regime.\cite{Jaime98,Broussard99,Zhao99x}
Crossovers from  $T^2$-like behavior to a flatter temperature dependence 
in the low-temperature region is widely seen.
Recently, resistivity data 
for narrow bandwidth compounds
including (Nd,Sr)MnO$_3$ and (Sm,Sr)MnO$_3$\cite{Akimoto00x} 
were shown to fit well with  $T^3$ at low temperatures.
We also note that a similar flattening in the low-temperature
resistivity is also observed 
in CrO$_2$, which is considered to be another half-metallic
system  \cite{Barry98,Li99}.

Let us discuss the applicability of the DMF.
In eq.~(\ref{defGdown}), we see that $G_\downarrow$
has nonzero $\Gamma_\downarrow$ for $k=k_{\rm F}$, $\omega=0$
in the limit $T\to0$. This behavior seems to be an artifact of
the DMF approximation,  which should be
recovered by taking into account proper vertex corrections
as a  magnon-drag phenomenon.
For the majority-band self-energy, however,
this recovery of coherence is not exhibited at  low temperatures.
The minority-band Green's function
should exhibit an incoherent behavior in the high-frequency region
$|\omega| \simge W^*$, irrespective of  the vertex correction.
The majority-band self-energy in  eq.~(\ref{defSigmaOneMagnon}) is
determined by the minority-band structure
in the frequency range $|\omega| \simle T$.
Since  $W^* \propto T^{3/2}$,
contributions from the
incoherent part of $G_\downarrow$ 
become dominant in eq.~(\ref{defSigmaOneMagnon}) when $W^* \ll T$.
Incoherent treatments for the minority-band Green's function
appear to be  valid at low temperatures.

Once  an incoherent minority band with the weight 
$z_\downarrow \propto \delta m$ is created, we have $\rho\propto T^3$.
Note that localization effects as well as strong electron-electron
 interactions enhance the incoherence of the minority band,
which enlarge the $T^3$ scaling region.
The crossover temperature $T^*$ is roughly estimated by
\begin{equation}
 T^*   \sim W^* \simeq ({0.06 W}/{2M(0)}) \cdot ({T^*}/{\Ds})^{3/2}.
\end{equation}
Using the parameters for (La,Sr)MnO$_3$,
$W\sim1{\rm eV}$, $\Ds \sim 10\ {\rm meV}$ and
$M(0)\sim 2$,
we obtain $T^* \sim 50 \ {\rm K}$, which is consistent with
the actual crossover temperature in experiments, 
$T^*_{\rm exp} \sim 30\ {\rm K}$,
where deviations from the $T^3$ behavior  emerge.
Above $T^*$, the minority band quasiparticles gain coherence
since $W^* > T$. The change in the electronic structure 
causes a deviation from the $T^3$ behavior.
Electronic properties at $T>T^*$, 
particularly  the  minority-band Green's function structure
and its influences to the 1MS,  remain for future study
in order to explain the $T^2$-like resistivity in experiments.

In the low-temperature region of manganites
$T\ll T_{\rm c}$, where
spin moments are considered to be almost saturated,
the roles of orbital degeneracies have
been emphasized \cite{Ishihara97,Imada98,Kilian98,Horsch99},
which should contribute to the Fermi-liquid-type $T^2$ resistivity
down to $T\to0$.
However, the $T^3$ behavior of the resistivity
at $T \simle T^*$ in Fig.~\ref{FigT3} indicates that
 the $T^2$ contributions are negligible.
Hence, concerning this temperature region of measurement,
$T \simle T^* \ll T_{\rm c}$,
the spin fluctuation scattering is much stronger than
other scattering mechanisms.

To summarize, the low-temperature resistivity of half-metals is investigated.
On the basis of
 the unconventional 1MS contribution, we obtain $T^3$ behavior
below the crossover temperature $T^*$.
We conclude that this is a unique character of  half-metals,
and may be used as a crucial test to make the distinction
from conventional itinerant ferromagnets.
Comparison with experimental data for
a possible half-metal (La,Sr)MnO$_3$
reveals  good agreement.
The author thanks A. Asamitsu and Y. Tokura
for providing the data in ref.~\citen{Urushibara95}
as well as for useful comments. He is also grateful
M.~Salamon, M.~Jaime, A.~J.~Millis, A. Gupta, Y. Moritomo 
and G.~M.~Zhao for valuable discussions.

%%%%%%   %%%%%   %%%%%   %%%%%   %%%%%   %%%%%   %%%%%   %%%%%   %%%%%   
%\bibliographystyle{jpsj}
%\bibliography{all,local}

%%%%%   %%%%%   %%%%%   %%%%%   %%%%%   %%%%%   %%%%%   %%%%%   %%%%%   

\end{document}